\documentclass[aps,prl,reprint]{revtex4-1}
\usepackage{graphicx}
\usepackage{epsfig}
\usepackage{bm}
\usepackage{color}
\usepackage{float}
\usepackage{dcolumn}
\usepackage{multirow} 
\usepackage{hyperref}
\usepackage{amsmath}

\graphicspath{{.}{../}}

\newcommand{\beq}{\begin{equation}}
\newcommand{\eeq}{\end{equation}}
\newcommand{\beqa}{\begin{eqnarray}}
\newcommand{\eeqa}{\end{eqnarray}}

\begin{document}

\title{A new and practical formulation for overlaps of Bogoliubov vacua}

\author{B. G. Carlsson}

\affiliation{Mathematical Physics, Lund University, S-221 00 Lund, Sweden}

\author{J. Rotureau}

\affiliation{Mathematical Physics, Lund University, S-221 00 Lund, Sweden}
\begin{abstract}
In this letter we present a new expression for the overlaps of wavefunctions
in Hartree-Fock-Bogoliubov based theories. Starting from the Pfaffian
formula by Bertsch \textit{et al} \citep{bertsch2012}, an exact and
computationally stable formula for overlaps is derived. We illustrate
the convenience of this new formulation with a numerical application
in the context of the particle-number projection method. This new
formula allows for substantially increased precision and versatility
in chemical, atomic, and nuclear physics applications, particularly
for methods dealing with superfluidity, symmetry restoration and uses
of non-orthogonal many-body basis states. 
\end{abstract}
\maketitle
\textit{Introduction.} A very successful approach in the context of
many-body theory is to incorporate correlation effects through symmetry
breaking followed by restoration of symmetries \citep{Bender2003}.
The intuitive picture gained from the symmetry-broken solutions has
provided much insight into the symmetry and beauty of nature. 
For instance, many nuclei can be described as intrinsically deformed
where both axial and reflection symmetry are broken \citep{Moller2006}.
Moreover, combining shape degrees of freedom with angular momentum
orientation, additional symmetries arise that can also be broken with
spectacular manifestations in nuclear spectra \citep{Frauendorf2001}.
For each broken symmetry new collective modes emerge, which complete
a picture of nuclear structure and nuclear spectra with an increasing
precision. As starting points, mean-field theories play a large role
because of their ability to find approximate solutions to the many-body
Schr{\"o}dinger equation at a relatively low computational cost \citep{rin80}.
A typical example is the Hartree-Fock-Bogoliubov (HFB) method where
the wavefunction solution is written as a product of independent quasiparticles.
Nuclear Physics applications of HFB coupled with symmetry-restoration
methods have allowed to accurately describe a vast swath of nuclear
properties such as binding energies, mean-square radii, deformation
and spectra. Under certain circumstances, such as for instance in
the case of shape coexistence, it becomes necessary to include additional
correlations between the quasiparticles. Correlations beyond the mean-field
can be included, for example, using the framework of the Generating
Coordinate Method (GCM) \citep{rin80,Bender2008,Yao2020,Malli,Trsic,Orest, FLGCM} where the
wavefunction is written as a linear combination of different non-orthogonal
mean-field configurations. In practice, the solution of GCM equations
and the application of symmetry-restoration methods require the precise
computation of overlap functions. The modulus of the overlap between
two HFB vacua can be computed with the Onishi formula \citep{Onishi}
thus leaving ambiguity on its sign. Several studies have subsequently
been dedicated to compute the overlap using various methods \citep{neergaard,Haider,Donau,bender,hara,oi,rodri,Bally2018,Mizusaki2018,Avez2012,schmid}.
In \citep{robledo2009}, Robledo derived an expression for the overlap
in terms of a Pfaffian. In a later publication \citep{bertsch2012},
Bertsch and Robledo extended the Pfaffian formulation to overlaps
between odd-A systems and overlaps for operators involved in symmetry
restoration methods. Let us consider two HFB vacua $|\Phi\rangle$
and $|\Phi'\rangle$ with even number parity (for an even-$A$ nucleus)
and the associated overlap ${\cal O}\equiv\langle\Phi|\Phi'\rangle$.
The formula by Bertsch and Robledo \citep{bertsch2012} gives ${\cal O}$
as: 
\begin{eqnarray}
{\cal {O}}=(-1)^{N/2}\frac{(detC)^{*}detC'}{{\displaystyle \prod_{i,i'}v_{i}v_{i'}'}}{\rm {pf}}\begin{pmatrix}V^{T}U & V^{T}V'^{*}\\
-V'^{\dagger}V & U'^{\dagger}V'^{*}
\end{pmatrix}\label{over_bertsch}
\end{eqnarray}
where $N$ is the (even) number of single-particle basis states 
and $U,V$ ($U',V'$) are matrices of the Bogoliubov transformation
associated with $|\Phi\rangle$ ($|\Phi'\rangle$) \citep{rin80}.
Using the Bloch-Messiah decomposition \citep{Bloch62}, one can write
$U=D\bar{U}C$ and $V=D^{*}\bar{V}C$ where $D$ and $C$ are both
unitary matrices and $D$ defines the so-called canonical basis associated
with $|\Phi\rangle$. $\bar{U}$ and $\bar{V}$ can be chosen as diagonal
and skew-symmetric, respectively. The matrix $\bar{V}$ is written
in terms of $N/2$ blocks of dimension $2\times2$ with elements ($v_{i},-v_{i}$)
where $v_{i}^{2}$ is the occupation probability of the canonical
basis state $i$ (the matrix elements $u_{i}$ of $\bar{U}$ are such
that $u_{i}^{2}+v_{i}^{2}=1$). We adopt the usual phase convention
$u_{i}>0$ and $v_{i}>0$ \citep{rin80} . For $v_{i}=1$, the level
$i$ is fully occupied whereas it is empty for $v_{i}=0$. $v'_{i}$
and $C'$ in (\ref{over_bertsch}) are obtained by the Bloch Messiah
decomposition of ($U',V'$). In practice, due to the tiny values of
$v_{i},v'_{i}$ for the least occupied levels, the computation of
the overlap (\ref{over_bertsch}) can become unstable. Indeed, in
this context, both the denominator ${\displaystyle \prod_{i,i'}v_{i}v_{i'}'}$
and the Pfaffian in the numerator of Eq.~(\ref{over_bertsch}) have
small values that can become out of the scope of the double-precision
data type. A potential solution to this issue could be to discard
levels for which $v_{i},v'_{i}\sim0$ in the computation of the overlap.
But in practice, it can become difficult to check the reliability
of such an approach. Indeed, one would need to check that the omitted
levels have a negligible contribution in the value of the overlap
and this would be done by increasing the number of levels considered.
Eventually one might again run into instability of the numerical computation.
 It is especially important to obtain precise values of overlaps
in the context of GCM where the basis states are not orthogonal and
requires the solution of a generalized eigenvalue problem in the overcomplete
basis. Other solutions to this issue have been proposed in \citep{gao2014}
where  small values of $v_{i}$ are replaced by an \textit{ad hoc}
tiny numerical parameter $\epsilon$. However, it would be advantageous
and more convenient for systematic calculations to be able to bypass
the introduction of such a parameter. It should be mentioned that
similar issues with multiplications of large and small terms were
also discussed in \citep{neergaard} where, using a different formulation
of the overlap, difficulties arise instead for $u_{i},u_{i'}'\sim0$.
This formulation was further generalized in \citep{Avez2012,robledo2011}
where tailored algorithmic procedures were introduced to allow for
inclusion of states having exactly $u_{i},u_{i'}'=0$. Nevertheless,
in practice the presence of states with very small but non-zero values
of $u_{i},u_{i'}'$ still requires the introduction of a cutoff. 

In this letter, we present a new and practical formulation of the
overlap (see Eq.~(\ref{final_over-1}) in the case of an even-$A$
system and Eq.~(\ref{odd7}) for an odd-$A$ system) which allows
for a precise and stable computation and is also amenable to controllable
truncations. 
This letter is organized as follows. We first show the main steps
involved in the derivation of the formula for the overlap between
two HFB vacua for an even-$A$ system and then show the expression
for the overlap in the case of an odd-$A$ system. We then illustrate
the numerical stability offered by this formulation by computing,
as a function of the number of canonical basis states included, the
matrix element of the particle-number projection operator. \\

We start the derivations from Eq. \ref{over_bertsch}. First, we want
to point out that this equation can not be used directly when including
empty levels since in that case, the expression involves products
of zero and infinity. 
The equation is however generally valid if one assumes a tiny occupation
$\epsilon$ for the empty levels. In our final expression Eqs.~(\ref{final_over-1},\ref{odd7})
these factors cancel out analytically and one may safely evaluate
the expression in the limit $\epsilon\rightarrow0$.

We denote $M$ the matrix argument of the Pfaffian in Eq.~(\ref{over_bertsch}).
Using the Bloch-Messiah decomposition for $U,V$ and $U',V'$ we obtain:
\begin{eqnarray}
 &  & M=\begin{pmatrix}-C^{T}\bar{V}\bar{U}C & -C^{T}\bar{V}D^{\dagger}D'\bar{V'}C'^{*}\\
C'^{\dagger}\bar{V'}D'^{T}D^{*}\bar{V}C & C'^{\dagger}\bar{U'}\bar{V'}C'^{*}
\end{pmatrix}\nonumber \\
 &  & =\begin{pmatrix}C^{T} & 0\\
0 & C'^{\dagger}
\end{pmatrix}\begin{pmatrix}-\bar{V}\bar{U} & -\bar{V}D^{\dagger}D'\bar{V'}\\
\bar{V'}D'^{T}D^{*}\bar{V} & \bar{U'}\bar{V'}
\end{pmatrix}\begin{pmatrix}C & 0\\
0 & C'^{*}
\end{pmatrix}\nonumber \\ \label{over3}
\end{eqnarray}
From Eq.~(\ref{over3}) and the relation 
\begin{eqnarray}
{\rm {pf}}(ABA^{T})={\rm det}(A){\rm {pf}}(B),\label{pfaf}
\end{eqnarray}
we can write: 
\begin{eqnarray}
{\rm pf}(M)={\rm det}C{\rm det}C'^{\dagger}{\rm {pf}}\begin{pmatrix}-\bar{V}\bar{U} & -\bar{V}D^{\dagger}D'\bar{V'}\\
\bar{V'}D'^{T}D^{*}\bar{V} & \bar{U'}\bar{V'}
\end{pmatrix}~~~~~\label{over4}
\end{eqnarray}
In order to avoid the numerical instability that arises in the computation
of the overlap directly from Eq.~(\ref{over_bertsch}) we factorize
the norms of the HFB vacua out of ${\rm pf}(M)$. This is achieved
in the following steps by first writing the matrix argument of the
Pfaffian on the right-hand side of Eq.~(\ref{over4}) as: 
\begin{equation}
\begin{pmatrix}\bar{V} & 0\\
0 & \bar{V'}
\end{pmatrix}\begin{pmatrix}\bar{U}\bar{V}^{-1} & D^{\dagger}D'\\
-D'^{T}D^{*} & -\bar{V'}^{-1}\bar{U'}
\end{pmatrix}\begin{pmatrix}-\bar{V} & 0\\
0 & -\bar{V'}
\end{pmatrix}\label{over5}
\end{equation}
Introducing the diagonal matrix 
\begin{equation}
\Lambda=\begin{pmatrix}\sqrt{v_{0}}\\
 & \sqrt{v_{0}}\\
 &  & \sqrt{v_{1}}\\
 &  &  & \ddots\\
 &  &  &  & \sqrt{v_{N/2-1}}
\end{pmatrix}
\label{lmat}
\end{equation}
and a similar matrix $\Lambda'$ with the occupation number $v_{i}'$,
we can rewrite (\ref{over5}) as : \begin{widetext} 
\begin{eqnarray}
\begin{pmatrix}\bar{V}\Lambda^{-1} & 0\\
0 & \bar{V'}\Lambda'^{-1}
\end{pmatrix}\begin{pmatrix}\Lambda\bar{U}\bar{V}^{-1}\Lambda & \Lambda D^{\dagger}D'\Lambda'\\
-\Lambda'D'^{T}D^{*}\Lambda & -\Lambda'\bar{V'}^{-1}\bar{U'}\Lambda'
\end{pmatrix}\begin{pmatrix}-\Lambda^{-1}\bar{V} & 0\\
0 & -\Lambda'^{-1}\bar{V'}
\end{pmatrix}\label{over6}
\end{eqnarray}
From the expression above, we can now write Eq.~(\ref{over4}) as:
\begin{eqnarray}
{\rm pf}(M) & = & {\rm det}C{\rm det}C'^{\dagger}{\rm pf}\left(\begin{pmatrix}\bar{V}\Lambda^{-1} & 0\\
0 & \bar{V'}\Lambda'^{-1}
\end{pmatrix}\begin{pmatrix}\Lambda\bar{U}\bar{V}^{-1}\Lambda & \Lambda D^{\dagger}D'\Lambda'\\
-\Lambda'D'^{T}D^{*}\Lambda & -\Lambda'\bar{V'}^{-1}\bar{U'}\Lambda'
\end{pmatrix}\begin{pmatrix}-\Lambda^{-1}\bar{V} & 0\\
0 & -\Lambda'^{-1}\bar{V'}
\end{pmatrix}\right)\label{over7}\\
 & = & {\rm det}C{\rm det}C'^{\dagger}{\rm det}(\bar{V}\Lambda^{-1}){\rm det}(\bar{V'}\Lambda'^{-1}){\rm {pf}}\begin{pmatrix}\Lambda\bar{U}\bar{V}^{-1}\Lambda & \Lambda D^{\dagger}D'\Lambda'\\
-\Lambda'D'^{T}D^{*}\Lambda & -\Lambda'\bar{V'}^{-1}\bar{U'}\Lambda'
\end{pmatrix}\label{over8}
\end{eqnarray}
\end{widetext} where Eq.~\ref{over8} has been obtained from Eq.~\ref{over7}
using once again the relation (\ref{pfaf}), 
 %
The matrix $\bar{V}\Lambda^{-1}$ is tridiagonal and skew-symmetric
with matrix elements $\sqrt{v_{i}}$ and $-\sqrt{v_{i}}$. As a consequence,
one has ${\rm det}(\bar{V}\Lambda^{-1})={\displaystyle \prod_{i}v_{i}}$
and similarly ${\rm det}(\bar{V'}\Lambda'^{-1})={\displaystyle \prod_{i}v'_{i}}$.
As it becomes clear now, the \textit{ad-hoc} introduction of $\Lambda$
and $\Lambda'$ has allowed to factorize the norm of the HFB vacua
out of the Pfaffian in Eq.~(\ref{over4}). Using Eq.~(\ref{over8})
in Eq.~(\ref{over_bertsch}), we now arrive at an expression of the
overlap as: 
\begin{align}
 & (-1)^{N/2}{\rm {pf}}\begin{pmatrix}\Lambda\bar{U}\bar{V}^{-1}\Lambda & \Lambda D^{\dagger}D'\Lambda'\\
-\Lambda'D'^{T}D^{*}\Lambda & -\Lambda'\bar{V'}^{-1}\bar{U'}\Lambda'
\end{pmatrix}\nonumber \\
= & (-1)^{N/2}{\rm {pf}}\begin{pmatrix}-\bar{U}\sigma & \Lambda D^{\dagger}D'\Lambda'\\
-\Lambda'D'^{T}D^{*}\Lambda & \sigma\bar{U'}
\end{pmatrix}\label{final_over-1}
\end{align}
where $\sigma$ is the $N\times N$ tridiagonal skew-symmetric matrix
with elements 1 and -1 \citep{suppl}. The expression (\ref{final_over-1})
is exact and allows for a stable numerical computation of overlaps
independently of how tiny the occupation number $v_{i}$ might be
(this includes the limit of vanishing occupations \textit{e.g.} in
the case of no pairing, where occupation numbers are exactly 0 for
levels with energy greater than the Fermi energy). Moreover, the formula
enables to systematically reduce the computing cost by including only
levels for which the occupation number $v_{i}$ ($v_{i}'$) is greater
than a given value $\eta$ and accordingly truncating the matrix in
Eq.~(\ref{final_over-1}). The quality of the truncation can then
be checked \textit{a posteriori} by decreasing $\eta$ to arbitrary
small values. Let us assume that for a given value of $\eta$, $n$
($n'$) canonical basis states fulfill the criteria $v_{i}\ge\eta$
($v'_{i}\ge\eta$). In that case, the values of the overlap in this
truncated space is given by \citep{suppl}: 
\begin{eqnarray}
(-1)^{n/2}{\rm pf}\begin{pmatrix}[-\bar{U}\sigma]_{n\times n} & [\Lambda D^{\dagger}D'\Lambda']_{n\times n'}\\
-[\Lambda'D'^{T}D^{*}\Lambda]_{n'\times n} & [\sigma\bar{U'}]_{n'\times n'}
\end{pmatrix}\label{over_trunc-2}
\end{eqnarray}

The extension for odd-$A$ systems is straightforward and the derivation
goes along the same line (see \citep{suppl}). Let us consider the
overlap between two odd-$A$ states written as one quasi-particle
creation operators acting on even-number-parity HFB vacua that is,
$\beta_{a}^{\dagger}|\Phi\rangle$ and $\beta_{a'}^{\dagger}|\Phi'\rangle$.
The corresponding overlap $\langle\Phi|\beta_{a}\beta_{a'}^{\dagger}|\Phi'\rangle$
is given as: 
\begin{eqnarray}
{\cal {O}}=(-1)^{N/2}{\rm {pf}}\begin{pmatrix}\mathcal{A} & \mathcal{B}\\
-\mathcal{B}^{T} & \mathcal{C}
\end{pmatrix}\label{odd7}
\end{eqnarray}
with matrix

\begin{equation}
\mathcal{A}=\left(\begin{array}{cc}
-\bar{U}\sigma & \Lambda D^{\dagger}D'\Lambda'\\
-\Lambda'D'^{T}D^{*}\Lambda & \sigma\bar{U'}
\end{array}\right),
\end{equation}
related to the even part of the system, 
\begin{align}
\mathcal{B} & =\left(\begin{array}{cc}
\Lambda D^{\dagger} & 0\\
0 & -\Lambda'D'^{T}
\end{array}\right)\left(\begin{array}{cc}
\mathbf{V}^{(a)*} & \mathbf{U}'^{(a')}\\
\mathbf{U}^{(a)*} & \mathbf{V}'^{(a')}
\end{array}\right),
\end{align}
related to the connection between the even and the odd part and 
\begin{equation}
\mathcal{C}=\left(\begin{array}{cc}
0 & \mathbf{U}^{(a)\dagger}\mathbf{U}'^{(a')}\\
-\mathbf{U}^{(a)\dagger}\mathbf{U}'^{(a')} & 0
\end{array}\right).
\end{equation}
related to the odd particles. The index $a$ ($a'$) in the equations
above is used to denote the column vectors associated with the two
blocked quasi-particles. As in the case of the overlap between even-number-parity
vacua, the expression for the overlap in the odd case is numerically
stable and allows for controllable truncations. In this context, the
truncated expression for the overlap can be written as \footnote{The matrix being skew symmetric, we do not write its lower triangle  part.}
(see \citep{suppl}): \begin{widetext} 
\begin{equation}
{\cal {O}}=(-1)^{n/2}{\rm {pf}}\begin{pmatrix}-[\bar{U}\sigma]_{n\times n} & [\Lambda D^{\dagger}D'\Lambda']_{n\times n'} & [\Lambda D^{\dagger}\mathbf{V}^{(a)*}]_{n\times1} & [\Lambda D^{\dagger}\mathbf{U}'^{(a')}]_{n\times1}\\
 & [\sigma\bar{U'}]_{n'\times n'} & -[\Lambda'D'^{T}\mathbf{U}^{(a)*}]_{n'\times1} & -[\Lambda'D'^{T}\mathbf{V}'^{(a')}]_{n'\times1}\\
 &  & 0 & [\mathbf{U}^{(a)\dagger}\mathbf{U}'^{(a')}]_{1\times1}\\
 &  &  & 0
\end{pmatrix}.
\end{equation}
\end{widetext}

\begin{figure}[htb]
\centering{}\includegraphics[clip,width=1\columnwidth]{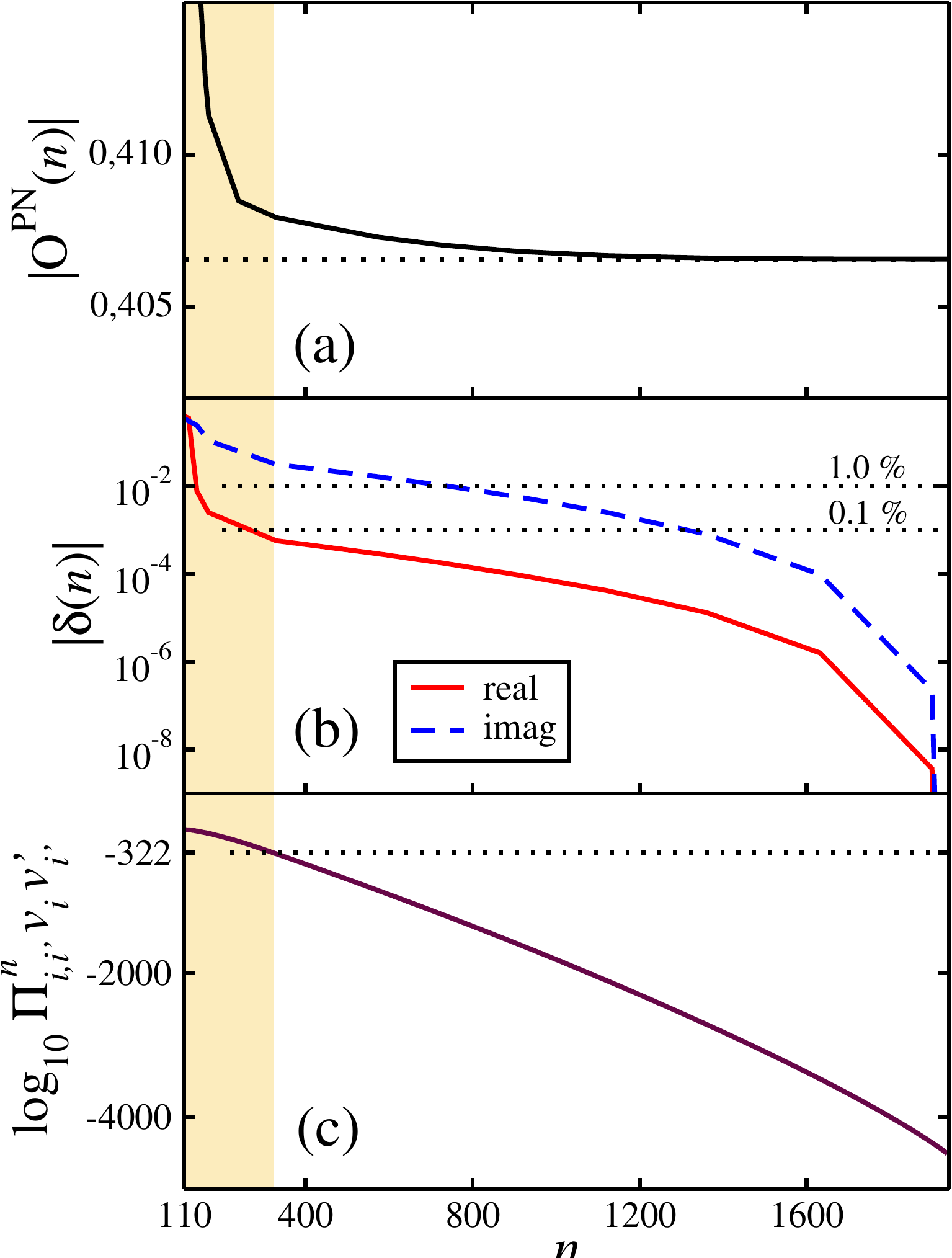}
\caption{Matrix element ${\cal {O}}^{{\rm PN}}(n)$ as a function of the number
of canonical basis states included in the computation $n$. Panel
(a) shows the modulus of the matrix elements at $\phi=\pi/10$ as
a function of $n$. The dotted-line corresponds to the value in the
full model space when $n=N$. Panel (b) shows the absolute value of the relative error $\delta(n)$
for both the real part (full line) and imaginary part (dashed-line)
of the matrix element (see text). The two horizontal dotted-lines
indicate the error at 1\% and 0.1\%. Panel (c) shows the decimal logarithm
of the denominator of the prefactor in Eq. (\ref{over_bertsch}). The horizontal dotted-line indicates the value $10^{-322}$, which corresponds approximately
to the smallest number that can be represented using the double-precision data type. The shadded area represents the range of validity of using Eq. (\ref{over_bertsch}) to  compute the overlap
at the  double-precision level.}
\label{fig1} 
\end{figure}

\textit{Example.} As a proof of principle, we now show results for
the computation of a matrix element involved in the particle-number
projection method for an even-$A$ nucleus. We denote this matrix
element ${\cal O}^{{\rm PN}}$: 
\begin{eqnarray}
{\cal {O}}^{{\rm PN}}\equiv\langle\Phi|e^{-i\hat{N}\phi}|\Phi\rangle=\langle\Phi|\Phi'\rangle\label{over_pn}
\end{eqnarray}
where $e^{-i\hat{N}\phi}$ is the gauge rotational operator for the
particle number projection, $\hat{N}$ the particle number operator
and $\phi$ the gauge angle. We focus on projecting the number of
neutrons and generate the occupation numbers with a HFB calculation
of $^{192}$Pb using a basis of 17 major oscillator shells, the SLY4
functional \citep{sly4} and a separable Gaussian pairing interaction
\citep{vesely12}. We rewrite ${\cal O}^{{\rm PN}}$ as the overlap
between the vacuum $|\Phi\rangle$ and a rotated vacuum $|\Phi'\rangle$
with the corresponding Bogoliubov-transformation matrices $U'=Ue^{i\phi}$
and $V'=Ve^{-i\phi}$. We set $\phi=\pi/10$ and for convenience,
we randomly generate the unitary matrices $D$ and $C$. We calculate
${\cal {O}}^{{\rm PN}}$ with Eq.~(\ref{over_trunc-2}) for different
values of the number $n$ of canonical basis states including in the
computation (we label ${\cal {O}}^{{\rm PN}}(n)$ the values obtained
when $n$ states are included). The exact value is obtained for $n=N=$1938
\textit{i.e.} when all single particle states have been included.
Results of the calculation are shown in Fig.~(\ref{fig1}). Panel
(a) shows the modulus of the matrix elements as a function of $n$
and as one can see, the value of the matrix elements smoothly converges
to the exact value as $n$ is increased. Panel (b) shows the modulus
of the real and imaginary part of the relative error defined as $\delta(n)=({\cal {O}}^{{\rm PN}}(n)-{\cal {O}}^{{\rm PN}}(N))/|{\cal {O}^{{\rm PN}}}(N)|$.
In order to have a relative error less than one percent around 800
states are needed. Panel (c) shows the decimal logarithm of $\prod_{i,i'}^{n}v_{i}v'_{i'}$
\textit{i.e.} the denominator of the prefactor of Eq.~(\ref{over_bertsch})
as a function of $n$. The horizontal dashed-line and arrow indicate
the value $10^{-322}$,
below which the computation of the overlap with Eq.~(\ref{over_bertsch})
is numerically unstable when computed with the double-precision data type. In such a case, the computation limit is reached
at $n=323$ for which $v_{i}=0.097593$.
Increasing the numerical precision would allow to extend the range of validity of using Eq. (\ref{over_bertsch}) to  compute the overlap albeit at the price of a higher computational burden. Since in the general case the product $\prod_{i,i'}^{n}v_{i}v'_{i'}$ may become arbitrary small, any calculation of Eq.~(\ref{over_bertsch}) with increased precision will eventually fail.

With the new formula all states can be included without encountering
any numerical instability. The ability to include all states is expected
to be particularly important for the description of the radial tail
of the wavefunction. A correct description of the wavefunction at
large distances is in turn critical for many applications involving
scattering and reaction processes \citep{calvin} such as alpha decay
\citep{Ward2013} or the computation of nucleon-nucleus optical potential
\citep{rot2,rot3,idi}.

\textit{Conclusion.} In this Letter we have presented a new, exact
and numerically stable formulation of overlap functions that appear
in beyond mean-field theories such as the GCM method. We have illustrated
the benefits offered by this formulation by computing a matrix element
for the particle-number projection method. We want to emphasize that
the derivations presented here are applicable to overlap functions
in the context of both even and odd number parity.

The expression
for the overlap can be combined with approaches to obtain the matrix
elements of the Hamiltonian for example by expressing these matrix
elements in a factorized form as sums of products of transition densities
multiplied by overlap functions \citep{rin80,Hu2014}. Such transition
densities are efficiently calculated in the canonical basis where
dimensions can be reduced by removing unoccupied states \citep{Yao2009,bonche,HEENEN0,VALOR,Tagami}.
The expressions derived in this letter may also be extended to cover cases when the
left and right vacua are expressed in different bases ( see e.g. \citep{Avez2012,robledo2011}).

The new formula
offers substantial improvements over current methods since exact values
can now be obtained even in large spaces without being limited by
the ability of representing very small or large numbers.
It adds no
extra computational effort or complexity and may be truncated in a
systematic way, which allows for a smooth convergence towards the
exact value. 
Finally, the new formula opens the door towards precision
calculations of nuclear reactions and nuclear structure where one
takes advantage of the power of symmetry breaking.\\

We acknowledge beneficial discussions with Andrea Idini. The authors
thank the Knut and Alice Wallenberg Foundation (KAW 2015.0021) and the Crawfoord foundation
for financial support.

\bibliography{biblio}

\end{document}